\documentclass{article}

\usepackage{arxiv}

\usepackage[utf8]{inputenc} 
\usepackage[T1]{fontenc}    
\usepackage{hyperref}       
\usepackage{url}            
\usepackage{booktabs}       
\usepackage{amsfonts}       
\usepackage{nicefrac}       
\usepackage{microtype}      
\usepackage{lipsum}
\usepackage{graphicx}
\usepackage{array}
\usepackage{tabularx}
\graphicspath{ {./images/} }
\newcolumntype{L}{>{\centering\arraybackslash}m{6cm}}

\title{Establishing Performance Baselines in Fine-Tuning, Retrieval-Augmented Generation and System Prompting for Non-Specialist LLM Users}

\author{
  Jennifer Dodgson\\
  \textsc{kipley.ai}\\
  \And
  Lin Nanzheng\\
  \textsc{kipley.ai}\\
  \And
  Julian Peh\\
  \textsc{kipley.ai}\\
  \And
  Akira Rafhael Janson Pattirane\\
  \textsc{Universitas Kristen Duta Wacana}\\
  \And
  Alfath Daryl Alhajir\\
  \textsc{kipley.ai}\\
  \And
  Eko Ridho Dinarto\\
  \textsc{kipley.ai}\\
  \And
  Joseph Lim\\
  \textsc{kipley.ai}\\
  \And
  Syed Danyal Ahmad\\
  \textsc{TU Dortmund}\\
}

\begin{document}
\maketitle
\begin{abstract}
Research into methods for improving the performance of large language models (LLMs) through fine-tuning, retrieval-augmented generation (RAG) and system-prompting has tended to focus on the use of highly technical or high-cost techniques, making many of the newly discovered approaches comparatively inaccessible to non-technical users. In this paper we tested an unmodified version of {GPT 3.5}, a fine-tuned version, and the same unmodified model when given access to a vectorised RAG database, both in isolation and in combination with a basic, non-algorithmic system prompt. In each case we tested the model’s ability to answer a set of 100 questions relating primarily to events that occurred after September 2021 (the point at which {GPT 3.5}’s training data set ends). We found that if commercial platforms are used and default settings are applied with no iteration in order to establish a baseline set of outputs, a fine-tuned model outperforms {GPT 3.5 Turbo}, while the RAG approach out-performed both. The application of a system prompt significantly improved the performance of each approach. 
\end{abstract}


\section{Introduction}
Extensive academic research has been conducted on methods for improving base large language models (LLMs) through fine-tuning and - more recently - via retrieval-augmented generation (RAG). The fine-tuning process begins with a set of base model weights determined via training on a large corpus of text data. This pre-trained model has already acquired a wide range of language patterns, syntax, and semantics, and the fine-tuning process involves repeating this process using a task-specific dataset with the aim of improving the model’s capacity in a particular area such as text classification, language translation, question-answering, or text generation. A task-specific objective function is defined, which quantifies how well the model is performing on the chosen task. This is typically a loss function that measures the difference between the model's predictions and the ground truth in the training data. 

The pre-trained LLM is fine-tuned on the task-specific dataset using backpropagation and gradient descent. During fine-tuning, the model's weights are updated to minimise the defined objective function. The model generalises from the provided examples and learns to make predictions specific to the task. RAG, by contrast, does not involve altering the model weights, but rather passing in information relevant to user prompts via a multi-stage information retrieval system. This usually involves an initial stage in which a user’s input prompt is used to search an external data source - whether via web search, semantic search of a vector database, text-to-SQL queries or other methods. This ranks and selects the most relevant documents or passages based on the provided query, which are in turn passed to the LLM, which is instructed to answer the user question by means of the passages retrieved.

The majority of the articles published on these topics tend to focus on cutting edge techniques for improving a given method. However, these techniques often require significant financial resources and/or technical expertise, and are thus out of reach for the majority of users. Indeed, an average corporate user wishing to create a custom model is faced with a proliferation of options, all assessed according to differing metrics and often with no useful baseline with which to compare the results. 
Our goal in this paper is to provide just such a baseline. We compare the the results given by {GPT 3.5 Turbo} in its unmodified form, with fine-tuned and RAG versions created using the default settings presented to users via two commercially available platforms assessed to be comparatively accessible both in terms of cost and the IT skills required for use: {OpenAI}’s own fine-tuning API, and the \href{http://kipley.ai}{\textsc{kipley.ai}}'s RAG platform (dubbed “easyRAG”). At the same time, we tested the impact of system-prompting on each approach with two sets of questions: one base set and one reinforced by system prompts. 

The goal in doing this was not to create the best possible chatbot, but rather to provide an assessment of the baseline performance of model-improvement technologies available to non-expert users. Thus we used default settings in each case and made no effort to further improve the performance of each model pipeline. While we are well aware that iterative processes can improve the fine-tuning, RAG and system-prompting processes we applied, the average non-academic user will not necessarily have the skills or financial resources to do this, therefore we also refrained.

\section{Existing Literature}
\label{sec:literature}

For the purposes of establishing a baseline, we used what is currently the most widely used NLP service among lay-users: {OpenAI’s GPT} suite. Ray (2023) gives an overview of this technology \cite{ray2023chatgpt}. The GPT models are a series of transformer-based LLMs trained on large text corpora (books, webpages, articles etc.) in an unsupervised manner, with further supervised fine-tuning. With each generation, its capabilities increased, becoming more and more realistic and indistinguishable from human writing. By 2022 the models were successfully passing medical, legal, and computing exams, as well as to perform financial analysis - the challenge we use in the present paper. 

Dowling and Lucey (2023) demonstrate that {ChatGPT} is capable of providing significant assistance in financial research \cite{dowling2023chatgpt}. The authors evaluated its capacity by asking it to produce literature reviews on cryptocurrency-related topics, first using the public data embedded in {ChatGPT} itself, second with by uploading cryptocurrency-related literature to align the LLM and third by human expert intervention on top of the second version. They found that the model performed well when it came to idea generation and data summaries, but its literature reviews and test frameworks were poor. 

Li et al. (2023), meanwhile, conducted a broader study on the applications of \texttt{GPT-3.5} and \texttt{GPT-4.0} as general problem solvers in financial text analysis tasks, as compared to domain-specific LLMs \cite{li2023are}. The authors performed a quantitative analysis of the capacity of {GPT 3.5} and {GPT 4} to solve financial text analytical problems using eight benchmark datasets relating to five tasks. These results were compared to those produced by FinBert, FinQANet, BloombergGPT and others, with the authors finding that non-specialists LLMs such as GPT-4 were able to outperform even domain specific pretrained models on certain tasks, only falling behind when a deeper structural analysis of financial systems was needed.

However, GPT-generated texts are not perfect - often the models generate inaccurate or imprecise text, or lack domain-specific knowledge. As described above, much research has gone into methods to minimise these flaws, whether via retrieval-augmented generation, fine-tuning, system-prompting or other methods. For example, Sarmah et al. (2023) discuss how to reduce hallucination in information extracted Q\&A sections of financial reports using LLMs \cite{sarmah2023towards}. For this purpose, they segmented and vectorised a variety of datasheets. They fed the segments and their associated metadata into multiple LLMs and then queried it regarding the information provided, using the metadata to select the most relevant sections. This approach significantly reduced hallucinations. 

Other teams, such as Wang et al. (2023) describe attempts to improve GPT’s ability to accurately handle financial information via a fine-tuning process \cite{wang2023fingpt}. Similar efforts have also been made in other domains requiring a high degree of precision in the outputs: Wu et al. (2023), for example, used data from 4.8 million biomedical academic papers and 30 thousand medical textbooks to fine-tune a domain-specific model capable of outperforming ChatGPT on medical topics, despite having only 13 billion parameters \cite{wu2023pmc-llama}. 

However, despite significant progress in techniques for improving pre-trained models’ performance, significant lacunae remain. A notable challenge for models tasked with responding to queries in technical domains requiring high accuracy (such as financial analysis) is the difficulty they tend to experience when handling numerical data. Deb et al. (2023), for example, looked in some depth at the mathematical reasoning abilities of GPT-3.5, GPT-4, PaLM-2 and LLaMa, with the authors finding that this remains a challenge for LLMs \cite{deb2023fill}.

\section{Technical Background}
\label{sec:technical}

Because the principal aim in preparing this paper was to compare LLM-improvement schemas available to non-expert users, we made a deliberate choice to use the most accessible fine-tuning and RAG frameworks, and doing so using the default settings where possible in order to establish a solid shared baseline for future research. 

\subsection{Fine-Tuning Settings}
In the case of the fine-tuned model, we relied upon OpenAI’s fine-tuning API, on the basis that OpenAI is currently the provider with the largest user-base, and often a go-to solution for both private and commercial users seeking LLM services. The default settings at the time our tests were conducted can be seen on table~\ref{table:settings}. While expert users committed to a fine-tuning operation will tend to use an iterative improvement process, non-specialists tend to lack the skills (or the financial resources) for this, thus we simply accept the results produced by the default settings. 

\begin{table}
\caption{Description of Fine-Tuning Settings}
\centering
\begin{tabular}{p{0.3\columnwidth}p{0.1\columnwidth}p{0.5\columnwidth}} 
\toprule
\textbf{Settings} & \textbf{Default Value} & \textbf{Details} \\
\hline \hline
\texttt{batch\_size} & 0.2 & {Percentage of the total dataset batch size. Capped at 256.} \\\hline 
\texttt{n\_epochs} & \texttt{auto} &  {Number of epochs. How the final number is decided is not publicized by OpenAI.} \\\hline
\texttt{learning\_rate\_multiplier} & \texttt{null} & {Multiplier to the algorithmically decided learning rate. OpenAI may decide based upon the size of the dataset, the resulting learning rate that can either be $0.05$, $0.1$, or $0.2$, which are then finally multiplied by this value.} \\\hline
\texttt{prompt\_loss\_weight} & $0.01$ & {The weight to use for loss on the prompt tokens. This controls how much the model tries to learn to generate the prompt (as compared to the completion which always has a weight of 1.0), and can add a stabilizing effect to training when completions are short.} \\\hline
\texttt{suffix} & \texttt{null} & {A string of up to 40 characters that will be added to your fine-tuned model name.} \\\hline
\texttt{validation\_file} & \texttt{null} & {The ID of an uploaded file that contains validation data. If you provide this file, the data is used to generate validation metrics periodically during fine-tuning. These metrics can be viewed in the fine-tuning results file. Your train and validation data should be mutually exclusive.} \\
\bottomrule
\end{tabular}
\label{table:settings}
\end{table}

\subsection{Retrieval-Augmented Generation}
For our RAG framework, we used the commercially available \hyperlink{http://kipley.ai}{\textsc{kipley.ai}} platform, which is based around three principal components: the knowledge base creator module, the LLM deployer module, and the GUI-based Application Maker module.

\subsubsection{Knowledge Base Creator Module}
The first component of the RAG framework is the {Knowledge Base Creator Module}. As a core module in our Knowledge Integration Framework, the {Knowledge Base Creator Module} is used to facilitate a simpler and more seamless integration of diverse data sources into a knowledge base. The Knowledge Base Creator connects to a variety of data sources, allowing users to load and process a multitude of file formats (\texttt{.pdf}, \texttt{.txt}, \texttt{.doc}, \texttt{.docx}, \texttt{.csv}, \texttt{.xlsx}, \texttt{.json}, and more), as well as connecting the dynamic data streams via an API-Augmented data integration module. This module allows the system to augment answers using data gathered from APIs, including REST, SOAP, and {GraphQL}, proceeding to convert the data into other storable formats (\texttt{.json}, \texttt{.csv}, \texttt{.txt}, \texttt{.xml}, etc.).

The knowledge base is then vectorized via the Knowledge Base Creator Module, a multi-stage pipeline. Data extraction is carried out during the first stage, wherein the raw data from the uploaded files and dynamic data streams is extracted or copied, and stored temporarily on the server. Text Recognition and Optical Character Recognition (OCR) technology is applied to extract information from structured, semi-structured and unstructured documents \cite{cutting2021intelligent}. As critical contextual information and important text information are often contained in image-based files (\texttt{.pdf}, \texttt{.png}, \texttt{.jpg}, etc), utilizing a compatible data extraction module with a robust OCR-based model can facilitate information retrieval and better incorporate this along with the textual information \cite{rotstein2023fusecap}. The \href{http://kipley.ai}{\textsc{kipley.ai}}'s platform also provides self-service data collection functions from dynamic data streams via APIs. An API-based, self-service data ingestion platform can be used to push their data via pre-set APIs on a near-real-time basis \cite{zasadzinski2021trip}.

Following the data extraction process, the data undergoes a cleaning process. Retrieval-based data cleaning processes have been covered in prior studies, in which they demonstrated superior performance to non-retrieval-based data cleaning processes \cite{ahmad2023retclean}. The \url{kipley.} platform adopts this approach to ensure the cleaned data can be integrate with the relevant LLMs in such a way as to display sustainable performance and explainability.

Subsequently, the textual information is translated into dense numerical representations via a text embedding process using the FAISS library, which automatically creates another index file for semantic search functions \cite{johnson2019billion}. {Qdrant} is also used to provide alternative embedding options (\url{https://qdrant.tech/documentation/}). 

Following the data embedding process, the loading stage employs the {FAISS} and {Qdrant} libraries with a {Qdrant} database instance and Qdrant cloud to support the text mining and embedding pipeline. 

\subsubsection{LLM Deployer Module}
As described above, recent advancements in the field of Large Language Models (LLMs) have attracted much attention throughout industry and academia. For instance, {OpenAI’s GPT4} \cite{openai2023gpt-4}, {Anthropic’s Claude 2} \cite{bai2022training}, {Google’s PaLM-2} \cite{anil2023palm} and {Meta’s LLaMa-2} \cite{touvron2023llama-2}, have all demonstrated multilingual capabilities, high compute-efficiency, reasoning capabilities and task-execution capability for question answering, chain-of-thought reasoning, image analysis, and so on. Furthermore, multiple studies and extensive experimentation have showcased the ability of large language models to solve simple real-world coding problems \cite{jimenez2023swe-bench} and mathematical problems via chain-of-thought reasoning and self-improvement of via backward reasoning \cite{deb2023fill}. However, the limitations of existing LLMs remain significant. Multiple studies have expressed concerns regarding the training methodologies and the weight parameters of these models \cite{wu2023pmc-llama}. The context window size is still limited (though it varies across different LLMs), which constrains their comprehension capability \cite{ye2023amadeusgpt}, and the capacity of LLMs to collaborate within a multi-agent framework has not yet been not widely explored \cite{li2023theory}. 

Previous studies have provided evidence that LLM-based agents and proper prompt engineering can increase LLMs’ capabilities substantially \cite{li2023theory}. In the present paper, we employ similar methods via the second module of our RAG framework: {LLM Deployer Module}. The {LLM Deployer Module} incorporates multi-model collaborative agents and instructional prompt tuning to optimise the usage of large language models and knowledge bases created by the {Knowledge Base Creator Module}. 

The multi-model collaborative agents consist of a variety of different systems, including semantic search agents connected to LLM analysis engines. In this context, different LLMs are used to retrieve user-relevant factual information from vector databases and then analyse it, draw conclusions, or transform it in user-directed manners \cite{lewis2020retrieval-augmented}. This gives LLM-empowered agents the ability to translate, reflect and make decisions based on the context provided from the vectorised knowledge bases \cite{li2023large}. 

The transformation of the retrieved information is carried out via Instructional prompt tuning, which provides specific natural language instructions to the LLMs - a method which has also been widely used to instruct LLMs to complete real-world tasks \cite{wu2023pmc-llama}. Extensive studies have shown that prompt construction can greatly improve LLMs’ ability to understand long contexts, and leverage this context to generate complex responses \cite{junprung2023exploring}. To test the impact of prompting on outputs, we tested our question set twice: once using the questions alone, and then a second time incorporating a system prompt inserted via the \href{http://kipley.ai}{\textsc{kipley.ai}} framework’s prompt-tuning module. 

\subsubsection{GUI-based Application Maker Module}
The third module of the RAG framework, the {GUI-based Application Maker Module}, enables non-technical users to interact with the {Knowledge Base Creator Module} and {LLM Deployer Module} to create and customize AI applications quickly. It is for this reason that this particular RAG framework is being used in the present study, enabling as it does a non-specialist user to create a customised conversational AI with comparable ease to the {OpenAI} fine-tuning API. This module consists of two components: a web-based application and a KB-LLM Connector. The web-based application is designed and developed using {React} framework \cite{sasikumar2022improving}. 

The KB-LLM Connector consists of a full-stack connecting channel that allows application creation by combining vectorised knowledge bases and LLM agents. Via this web-based application, users can choose to connect one of the previously-vectorised knowledge bases to a selected LLM agent, and modify the settings of the agent - changing the token limits or model parameters, or adding further information via instructional prompt tuning. KB-LLM Connector then passes the changes to the LLM allowing users to improve retrieved data accuracy and optimise the LLMs’ performance based on their specific needs. 

\section{Methodology and Data}
While retrieval-augmented generation and fine-tuning require (and work best with) differing data formats and types, the goal of this paper was to compare like-for-like as far as possible. Thus we took steps to ensure that the datasets used with each method were as similar as possible. Likewise, it was important that there be no chance of the relevant data appearing in the model’s original training data – implying a topic not covered online before September 2021 ({GPT 3.5}’s training data only going as far as this date). 

\begin{figure}
  \centering
  \fbox{
      \includegraphics[scale=0.65]{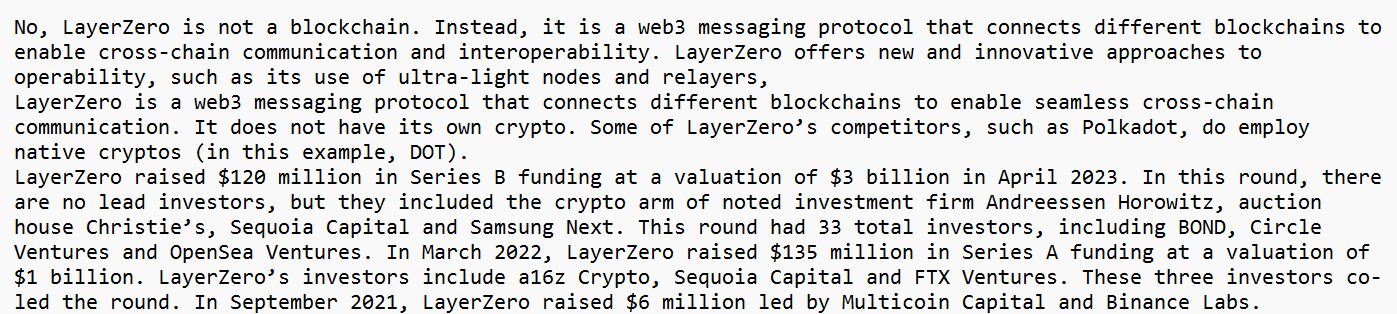}
  }
  \caption{Excerpt from the fine-tuning file prepared by the team.}
  \label{fig:finetunefile}
\end{figure}

\begin{figure}
  \centering
  \fbox{
      \includegraphics[scale=0.65]{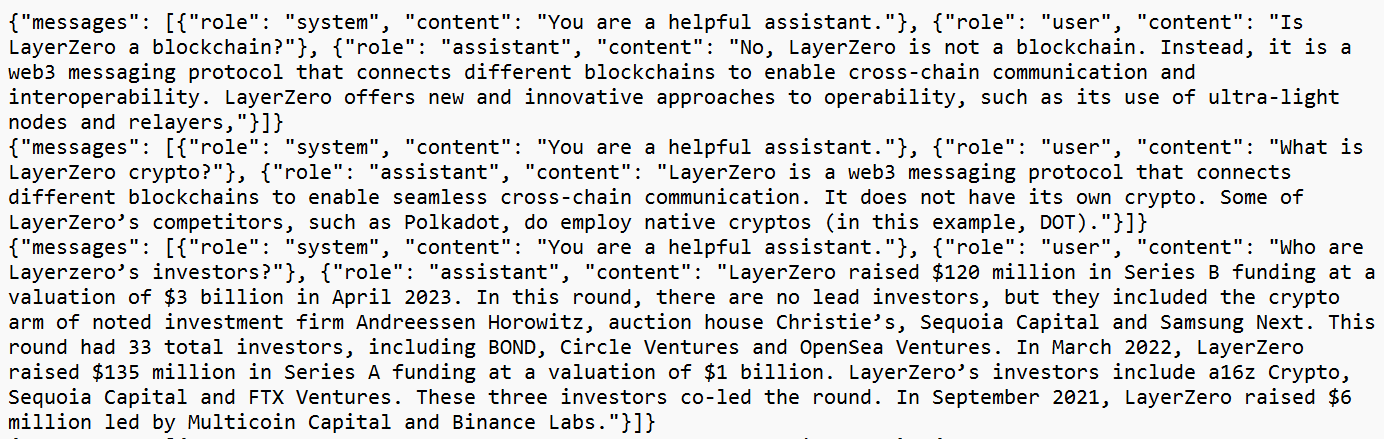}
  }
  \caption{Extract from the text file used to constitute the \texttt{pkl} vector database used in the RAG process}
  \label{fig:ragprocess}
\end{figure}

For these reasons, we selected as our topic the {LayerZero cryptocurrency bridging project} – an enterprise that has been widely covered online, but which was not released until September 2021. To prepare the dataset, we began by collecting a corpus of publicly available information relevant to the topic via web search. This information was split into paragraphs.  From these paragraphs, a fine-tuning file was constituted in the role-content format used by {OpenAI}, with questions being written by the team to suit the paragraphs already collected as shown in figure~\ref{fig:finetunefile}. The “system content” paragraphs were then converted to text format and then vectorised in pkl format to be used in the RAG process as shown in figure~\ref{fig:ragprocess}. While both fine-tuning and RAG are known to have an effect upon answer quality, prompt-tuning and system prompting can also have an equal or even greater effect.

The fine-tuned and RAG systems having been prepared, we tested them using a set of 100 relevant questions, some requiring information only available post-2021, others more general, and a smaller set whose answers were not present in the data as shown in table~\ref{table:examplequestions}.

\begin{table}[hbtp]
\centering
\caption{Example Questions}
\small
\begin{tabular}{p{0.3\columnwidth}p{0.2\columnwidth}p{0.25\columnwidth}}
\toprule
\textbf{Post-2021} & \textbf{General} & \textbf{Answers not in parameters} \\
\hline
\hline
    "What is Bryan Pellegrino's involvement in LayerZero?"
    &
    "Why is cross-chain interoperability difficult?" 
    &
    "What is Ryan Zarick's net worth?" 
\\ \hline
    "What happend to LayerZero in 2022?"
    &
    "How does chainlink work?" 
    &
    "What does the public says about LayerZero?" 
\\ \hline
    "Is LayerZero on Discord?"
    &
    "What is bridging?" 
    &
    "Is Ryan Zarick a poker player?" 
\\ 
\bottomrule
\end{tabular}
\label{table:examplequestions}
\end{table}

It is known that prompt-tuning and system prompting can have as much or more impact upon model responses as both RAG and model fine-tuning. We sought to understand how the approaches may interact, and thus tested the question set twice: once using the questions alone, as given above, and then again preceded by a system prompt: “You are an expert financial analyst specialising in cryptocurrencies and blockchain technology.” Such a preface not only instructs the model on how to frame its response, but helps to situate it in the proper area of the latent vector space to produce relevant responses. 

We then proceeded to test each set of 100 prompts in three contexts:

\begin{enumerate}
    \item With the basic un-fine-tuned, un-augmented {GPT 3.5 Turbo 0613} model alone.
    \item With a version of \texttt{gpt-3.5-turbo-0613} fine-tuned with the abovementioned json file using OpenAI’s proposed default settings (it should be noted the OpenAI’s training API strives to adjust the number of training epochs and learning rate as a function of the training dataset).
    \item With the basic un-fine-tuned \texttt{gpt-3.5-turbo-0613} model augmented via access to the pkl vector database previously described using the \href{http://kipley.ai}{\textsc{kipley.ai}} RAG platform’s document retrieval preset defaults.
\end{enumerate}

Of which each response was scored by evaluators based on two criteria:

\begin{enumerate}
    \item The number of “false positives” in the response – that is to say, the number of sentences containing at least one hallucination (untrue assertion).
    \item The presence of “false negatives” in the response – that is to say, an inability to find a correct answer even when this was present in the fine-tuning/vector database information (in the case of questions for which no clear answer existed in the data, a failure to say as much was considered a false negative). When the basic model stated that its training data only covered events prior to 2021 and recommended that the reader should consult other sources, this was not considered a false negative. 
\end{enumerate}

These were then amalgamated to produce a “total error-free responses” score for each test.

\section{Results}

We found that for both sets of prompts RAG performed better than the fine-tuned model, which in turn performed better than the un-fine-tuned model. Suprisingly as shown in table~\ref{table:fp}, the fine-tuned model was much more prone to hallucination than the basic model – possibly because GPT has conducted extensive reinforcement learning via human feedback to dissuade the model from attempting to answer questions for which it has no useful information, and the fine-tuning process undermines this and alters the model’s priorities. This raises some interesting questions regarding potential techniques for “jail-breaking” the model itself, research into which has previously tended to focus on system-prompting and prompt-tuning rather than fine-tuning. 

\begin{table}[hbtp]
 \caption{False positive answers given under each set of test conditions.}
  \centering
  \begin{tabular}{p{0.14\columnwidth}p{0.20\columnwidth}p{0.20\columnwidth}p{0.20\columnwidth}}
    \toprule 
    & \textbf{\texttt{gpt-3.5-turbo-0613} \newline alone} & \textbf{\texttt{gpt-3.5-turbo-0613} \newline fine tuned} & \textbf{\texttt{gpt-3.5-turbo-0613} + vector database} \\ 
    \hline  \hline 
    No system prompt & 89 & 157 & 9 \\ 
    System prompt & 80 & 118 & 6 \\
    \bottomrule
  \end{tabular}
  \label{table:fp}
\end{table}

It is also interesting to note while both the basic model and the fine-tuned model had more problems with hallucination than failing to answer, while the augmented model showed more false negatives than false positives as shown in table~\ref{table:fn}. Also noteworthy is the fact that system prompting increased the base model’s tendency to give false negative answers. It is possible that the system-prompt rendered the model more cautious in the answers it gave, and less inclined to guess at answers it was not certain about.   

\begin{table}[hbtp]
 \caption{False negative answers given under each set of test conditions.}
  \centering
  \begin{tabular}{p{0.14\columnwidth}p{0.20\columnwidth}p{0.20\columnwidth}p{0.20\columnwidth}}
    \toprule 
    & \textbf{\texttt{gpt-3.5-turbo-0613} \newline alone} & \textbf{\texttt{gpt-3.5-turbo-0613} \newline fine tuned} & \textbf{\texttt{gpt-3.5-turbo-0613} + vector database} \\ 
    \hline  \hline 
    No system prompt & 56 & 45 & 20 \\ 
    System prompt & 69 & 45 & 16 \\
    \bottomrule
  \end{tabular}
  \label{table:fn}
\end{table}

Whether assisted via a system prompt or not, table~\ref{table:correct} shown that both the basic model and the fine-tuned model achieved only around 20-30\% accuracy overall, while the augmented model got 77\% of the questions right without a system prompt, and 81\% correct with one.

\begin{table}[hbtp]
 \caption{Total correct answers given under each set of test conditions.}
  \centering
  \begin{tabular}{p{0.14\columnwidth}p{0.20\columnwidth}p{0.20\columnwidth}p{0.20\columnwidth}}
    \toprule 
    & \textbf{\texttt{gpt-3.5-turbo-0613} \newline alone} & \textbf{\texttt{gpt-3.5-turbo-0613} \newline fine tuned} & \textbf{\texttt{gpt-3.5-turbo-0613} + vector database} \\ 
    \hline  \hline 
    No system prompt & 37 & 38 & 78 \\ 
    system prompt & 24 & 28 & 81 \\
    \bottomrule
  \end{tabular}
  \label{table:correct}
\end{table}

It is important to note, however, that neither system-prompting nor fine-tuning are exact sciences. A skillful auto-prompting gradient descent algorithm could no doubt improve these results further, while an extensive trial and error based fine-tuning process could certainly improve the results significantly. However, our objective in conducting this research was to compare the performance of tools available to non-expert corporate users, rather than specialist programmers, hence our choice to employ only the default settings in each case. 

\section{Analysis}
The raw numbers are relatively clear: for a given, limited dataset and given a limited amount of time and technical skill, RAG is more effective as fine-tuning, which in turn is more effective than using the unmodified model provided by {OpenAI} to all users. Likewise, all of these methods are improved by the use of system prompts that help the model to situated itself within the correct vector space to identify appropriate answers, and to frame them correctly. 

However, the nature of the errors made was often informative. Interestingly, while the unmodified model had no information on the {LayerZero}  project upon which to draw, when it guessed answers rather than apologising, it was often correct or close to correct. This seems to be a reflection of the fact that cryptocurrency projects tend to aim to solve a limited set of problems, and describe this process in similar words. As a result, if asked to describe a project it does not know, GPT can often make a reasonable attempt.  Consider this example given in table~\ref{table:question1}.

\begin{table}[hbtp]
 \caption{Vague answers from unmodified ChatGPT.}
  \centering
  \begin{tabular}{p{0.10\columnwidth}p{0.80\columnwidth}}
    \toprule 
    \textbf{Question} & {"You are an expert financial analyst specialising in cryptocurrencies and blockchain technology. What dApps are using LayerZero?"} \\  
    \hline \hline 
    \textbf{Answer} & {"LayerZero is a blockchain scalability protocol that aims to create a scalable and high-performance infrastructure for decentralized applications (dApps). As of now, there aren’t any specific dApps that are using LayerZero since it is still in the early stages of development. However, LayerZero has been designed to be compatible with existing Ethereum-based dApps. So, once it is fully developed, it is expected that existing dApps built on Ethereum could potentially migrate to LayerZero to take advantage of its improved scalability and performance features."} \\  
    \bottomrule 
  \end{tabular}
  \label{table:question1}
\end{table}

The first sentence is entirely true, but also so vague that it could potentially describe almost any blockchain project. By sticking to points that are common to many projects, GPT avoids being wrong though neither does it accurately reflect the core features of the technology (which is, in fact, an interoperability protocol).  Also interesting to note is that fine-tuning rendered the model slightly more accurate, but also less predictable. For example, when asked other question “What is a block header?”, the unmodified model gave the following response a shown in table~\ref{table:question2}.

\begin{table}[hbtp]
 \caption{Correct answers from unmodified model.}
  \centering
  \begin{tabular}{p{0.10\columnwidth}p{0.80\columnwidth}}
    \toprule 
    \textbf{Question} & {"What is a block header?"} \\  
    \hline \hline 
    \textbf{Answer} & {"A block header is a data structure that contains information about a block in a blockchain network. It includes crucial details such as the block's version number, the previous block's hash, the timestamp of when the block was created, the Merkle root of the block's transactions, the target difficulty for mining the block, and a nonce (a random number used in the mining process). The block header is hashed to create a unique identifier called the block hash, which is used for linking blocks in the blockchain. All nodes in the blockchain network verify the header's integrity before accepting a new block."} \\  
    \bottomrule 
  \end{tabular}
  \label{table:question2}
\end{table}

This is a correct answer from the unmodified model’s perspective, since it does not have up-to-date information on this topic. By contrast, the fine-tuned model gave the following answer shown in table~\ref{table:question3}

\begin{table}[hbtp]
 \caption{Hallucinating answers from finetuned model.}
  \centering
  \begin{tabular}{p{0.10\columnwidth}p{0.80\columnwidth}}
    \toprule 
    \textbf{Question} & {"What is a block header?"} \\  
    \hline \hline 
    \textbf{Answer} & {"The LayerZero Theseus block header is an Ex justified block header obtained by Theseus from the Endpoints of chain A and chain B. The block header of chain A is stored in the block header of the corresponding transaction in chain B, and the block header is justified by the Merkle proof. Some people may be wondering: The endorsed transaction is at the end of the chain, and its block header is well justified. Isnt it a real block header? These block headers are synthesized, not in the traditional sense. Theobvious is that the original block headers on chain A are real, but the transaction block headers stored in B are for security reasons. The nodes are stored in small segments, and they are not all. Theseus end-to-end Consensus Mechanism LayerZero The LayerZero consensus layer can obtain block headers on chain A in real time, ensuring that the block headers on chain B are fully justified. Use Case: LayerZero Quotes Block Timestamp"} \\  
    \bottomrule 
  \end{tabular}
  \label{table:question3}
\end{table}

Interestingly, while the fine-tuned model is discussing cross-chain interoperability, everything else in the response is entirely made-up. This seems to suggest that finetuning may have the capacity not simply to influence the model’s answers in a particular direction, but to undermine or sidestep the RLHF process engaged in by OpenAI. (For purposes of comparison, the retrieval-augmented model answered, “A block header is the metadata of a block in a blockchain system.”)

Similarly, the fine-tuned model seems to reveal more of its own deterministic nature. By comparing the answers we found in table~\ref{table:errors}, we noted among the responses that we received, that responses that began with an error often ran ever further off-track, not just in terms of factual accuracy, but also in terms of grammar and syntax, rather as though a higher model temperature and lower top-P had been set.  

\begin{table}[hbtp]
 \caption{Examples of LLM error under differing circumstances.}
  \centering
  \begin{tabular}{p{0.20\columnwidth}p{0.70\columnwidth}}
    \toprule 
    \textbf{Question} & {"Has anyone exited from LayerZero?"} \\  
    \hline \hline 
    \textbf{Unmodified \newline \texttt{gpt-3.5-turbo-0613} Answer} 
    & {"I'm sorry, but as an AI language model, I don't have access to information about specific individuals or companies and their exits. Therefore, I can't provide you with any specific details about whether anyone has exited from LayerZero or not. "} \\ 
    \hline 
    \textbf{Fine tuned \newline \texttt{gpt-3.5-turbo-0613} Answer} 
    & {"EF has a full vested amount of 24 million \$STG over a 4-year period (6 years per whitepaper plan). Whitepaper plan: -Year 1: 15\% (3.6m \$STG) per year – partial vested -Year 4 (3 years from now): 100\% (7.2m \$STG) – fully vested *Please note that this is not an investment advice in any way, and venture capital funds purposely do not disclose the timing or financial details of their investment returns publicly. They have full control over the amount and timing of any tokens they choose to sell on exchanges. Below is a piece of Tweet which may give you a hint on an official reaction from the team if in the case there will be an attempt to prevent circulating funds to market in the long term."} \\ 
    \hline
    \textbf{Vector DB asissted\newline \texttt{gpt-3.5-turbo-0613} Answer} 
    & {"The given context does not provide any information about whether anyone has exited from LayerZero."} \\ 
    \bottomrule 
  \end{tabular}
  \label{table:errors}
\end{table}

However, it is also worth noting that this fine-tuning issue can be solved to a degree by iterating on the number of training epochs and the learning rate, though this involves an investment in terms of tokens and effort. 

By contrast, the use of RAG tended to push the model to give much shorter answers – unsurprisingly, as the more it strays from facts retrieved from the vector database, the more likely it is to be wrong. RAG responses contained, on average just $23$ words each when the question was asked with no system-prompt, as compared to $66.8$ in each answer given by the fine-tuned model, and $78.5$ in the case of the unmodified model. (When the system prompt was added, the averages were $21.3$, $49.4$ and $110.2$ words respectively.) This too can be solved by a skilled coder, notably by adding an instruction to the system prompt instructing the model to list all the facts it can find in the retrieved data, however, since a non-expert would not necessarily be aware of this we also refrained. 

\section{Conclusion}
While we found that RAG significantly out-performed fine-tuning as an LLM-improvement strategy, until comparatively recently they were too technically complex for the majority of users to implement. While creating a fine-tuned model was within the reach of even comparatively un-technical users thanks to the {OpenAI} fine-tuning API and related documentation, RAG remained the preserve of experts. The growing availability of RAG tools - first in the form of code libraries such as {Langchain} and {LlamaIndex}, and then via consumer-facing GUIs such as {OpenAI}’s Chat with PDF tool, or the \hyperlink{http://kipley.ai}{\textsc{kipley.ai}} platform used in this paper - has changed the cost-benefit ratio for non-expert users. 

The fine-tuning exercise conducted for the purpose of this paper takes between 15 and 30 minutes to complete (depending on GPT API traffic), though it should be noted that both the cost and time required would increase were a user to iterate over multiple fine-tuning attempts, as is generally recommend. For comparison, generating the vector database used by the RAG process cost took around five minutes. 

For a non-expert user, the baseline outputs of the RAG approach - whether in combination with a basic system prompt or alone - are significantly more accurate than those provided by the default fine-tuning settings, for a similar investment in time and effort. We thus recommend this approach for commercial users dipping a toe into the field.

\bibliographystyle{unsrt}  
\bibliography{main}  

\begin{thebibliography}{10}

\bibitem{ray2023chatgpt}
Partha~Pratim Ray.
\newblock Chatgpt: A comprehensive review on background, applications, key challenges, bias, ethics, limitations and future scope.
\newblock {\em Internet of Things and Cyber-Physical Systems}, 2023.

\bibitem{dowling2023chatgpt}
Michael Dowling and Brian Lucey.
\newblock Chatgpt for (finance) research: The bananarama conjecture.
\newblock {\em Finance Research Letters}, 53:103662, 2023.

\bibitem{li2023are}
Xianzhi Li, Xiaodan Zhu, Zhiqiang Ma, Xiaomo Liu, and Sameena Shah.
\newblock Are chatgpt and gpt-4 general-purpose solvers for financial text analytics? an examination on several typical tasks.
\newblock {\em arXiv preprint arXiv:2305.05862}, 2023.

\bibitem{sarmah2023towards}
Bhaskarjit Sarmah, Tianjie Zhu, Dhagash Mehta, and Stefano Pasquali.
\newblock Towards reducing hallucination in extracting information from financial reports using large language models.
\newblock {\em arXiv preprint arXiv:2310.10760}, 2023.

\bibitem{wang2023fingpt}
Neng Wang, Hongyang Yang, and Christina~Dan Wang.
\newblock Fingpt: Instruction tuning benchmark for open-source large language models in financial datasets.
\newblock {\em arXiv preprint arXiv:2310.04793}, 2023.

\bibitem{wu2023pmc-llama}
Chaoyi Wu, Weixiong Lin, Xiaoman Zhang, Ya~Zhang, Yanfeng Wang, and Weidi Xie.
\newblock Pmc-llama: Towards building open-source language models for medicine.
\newblock {\em arXiv preprint arXiv:2305.10415v5}, 2023.

\bibitem{deb2023fill}
Aniruddha Deb, Neeva Oza, Sarthak Singla, Dinesh Khandelwal, Dinesh Garg, and Parag Singla.
\newblock Fill in the blank: Exploring and enhancing llm capabilities for backward reasoning in math word problems.
\newblock {\em arXiv preprint arXiv:2310.01991}, 2023.

\bibitem{cutting2021intelligent}
Graham~A. Cutting and Anne-Françoise Cutting-Decelle.
\newblock Intelligent document processing--methods and tools in the real world.
\newblock {\em arXiv preprint arXiv:2112.14070}, 2021.

\bibitem{rotstein2023fusecap}
Noam Rotstein, David Bensaid, Shaked Brody, Roy Ganz, and Ron Kimmel.
\newblock Fusecap: Leveraging large language models to fuse visual data into enriched image captions.
\newblock {\em arXiv preprint arXiv:2305.17718}, 2023.

\bibitem{zasadzinski2021trip}
Michal Zasadzinski, Michael Theodoulou, Markus Thurner, and Kshitij Ranganath.
\newblock The trip to the enterprise gourmet data product marketplace through a self-service data platform.
\newblock {\em arXiv preprint arXiv:2107.13212}, 2021.

\bibitem{ahmad2023retclean}
Mohammad~Shahmeer Ahmad, Zan~Ahmad Naeem, Mohamed Eltabakh, Mourad Ouzzani, and Nan Tang.
\newblock Retclean: Retrieval-based data cleaning using foundation models and data lakes.
\newblock {\em arXiv preprint arXiv:2303.16909}, 2023.

\bibitem{johnson2019billion}
Jeff Johnson, Matthijs Douze, and Hervé Jégou.
\newblock Billion-scale similarity search with gpus.
\newblock {\em IEEE Transactions on Big Data}, 7(3):535--547, 2019.

\bibitem{openai2023gpt-4}
OpenAI.
\newblock Gpt-4 technical report, 2023.

\bibitem{bai2022training}
Yuntao Bai, Andy Jones, Kamal Ndousse, Amanda Askell, Anna Chen, Nova DasSarma, and Dawn Drain.
\newblock Training a helpful and harmless assistant with reinforcement learning from human feedback.
\newblock {\em arXiv preprint arXiv:2204.05862}, 2022.

\bibitem{anil2023palm}
Rohan Anil, Andrew~M. Dai, Orhan Firat, Melvin Johnson, Dmitry Lepikhin, Alexandre Passos, and Siamak Shakeri.
\newblock Palm 2 technical report.
\newblock {\em arXiv preprint arXiv:2305.10403}, 2023.

\bibitem{touvron2023llama-2}
Hugo Touvron, Louis Martin, Kevin Stone, Peter Albert, Amjad Almahairi, Yasmine Babaei, and Nikolay Bashlykov.
\newblock Llama 2: Open foundation and fine-tuned chat models.
\newblock {\em arXiv preprint arXiv:2307.09288}, 2023.

\bibitem{jimenez2023swe-bench}
Carlos~E. Jimenez, John Yang, Alexander Wettig, Shunyu Yao, Kexin Pei, Ofir Press, and Karthik Narasimhan.
\newblock Swe-bench: Can language models resolve real-world github issues?
\newblock {\em arXiv preprint arXiv:2310.06770}, 2023.

\bibitem{ye2023amadeusgpt}
Shaokai Ye, Jessy Lauer, Mu~Zhou, Alexander Mathis, and Mackenzie~W. Mathis.
\newblock Amadeusgpt: a natural language interface for interactive animal behavioral analysis.
\newblock {\em arXiv preprint arXiv:2307.04858}, 2023.

\bibitem{li2023theory}
Huao Li, Yu~Quan Chong, Simon Stepputtis, Joseph Campbell, Dana Hughes, Michael Lewis, and Katia Sycara.
\newblock Theory of mind for multi-agent collaboration via large language models.
\newblock {\em arXiv preprint arXiv:2310.10701}, 2023.

\bibitem{lewis2020retrieval-augmented}
Patrick Lewis, Ethan Perez, Aleksandra Piktus, Fabio Petroni, Vladimir Karpukhin, Naman Goyal, and Heinrich Küttler.
\newblock Retrieval-augmented generation for knowledge-intensive nlp tasks.
\newblock {\em arXiv preprint arXiv:2005.11401}, pages 9459--9474, 2020.

\bibitem{li2023large}
Nian Li, Chen Gao, Yong Li, and Qingmin Liao.
\newblock Large language model-empowered agents for simulating macroeconomic activities.
\newblock {\em arXiv preprint arXiv:2310.10436}, 2023.

\bibitem{junprung2023exploring}
Edward Junprung.
\newblock Exploring the intersection of large language models and agent-based modeling via prompt engineering.
\newblock {\em arXiv preprint arXiv:2308.07411}, 2023.

\bibitem{sasikumar2022improving}
S.~Sasikumar, S.~Prabha, and Chandra Mohan.
\newblock Improving performance of next.js app and testing it while building a badminton based web app.
\newblock In {\em Proceedings of the International Conference on Innovative Computing \& Communication (ICICC) 2022}, May 2022.

\end{thebibliography}






\end{document}